\documentclass[aps,prb,twocolumn,superscriptaddress]{revtex4}

\pdfoutput=1

\usepackage{graphicx}
\usepackage{dcolumn}
\usepackage{float}
\usepackage{amssymb}
\usepackage{xcolor}

\begin{document}

\title{Interface limited growth of heterogeneously
       nucleated ice in supercooled water}

\author{Razvan A. Nistor}
\affiliation{Department of Chemistry, Columbia University, 
             3000 Broadway, MC 3103, New York, NY 10027}
\author{Thomas E. Markland}
\affiliation{Department of Chemistry, Stanford University,
             333 Campus Drive, Stanford, California 94305}
\author{B. J. Berne} \email{bb8@columbia.edu}
\affiliation{Department of Chemistry, Columbia University,
             3000 Broadway, MC 3103, New York, NY 10027}

\begin{abstract}
Heterogeneous ice growth exhibits a maximum in freezing rate arising 
from the competition between kinetics and the thermodynamic driving
force between the solid and liquid states.  Here, we use molecular 
dynamics simulations to elucidate the atomistic details of this competition,
focusing on water properties in the interfacial region along the secondary 
prismatic direction.  The crystal growth velocity is maximized when the
efficiency of converting interfacial water molecules to ice, collectively
known as the attachment kinetics, is greatest.  We find water molecules that 
contact the intermediate ice layer in concave regions along the atomistically 
roughened surface are more likely to freeze directly.  An increased roughening of
the solid surface at large undercoolings consequently plays an important
limiting role on the rate of ice growth, as water molecules are unable to 
integrate into increasingly deeper surface pockets.
These results provide insights into the molecular mechanisms for self-assembly
of solid phases that are important in many biological and
atmospheric processes. 
\end{abstract}

\maketitle

\section{Introduction} \label{sec:intro}

Self-assembly of a disordered liquid to an ordered solid is one of the most
basic physical processes that occurs in
nature.~\cite{kirkpatrick75mineral,kotrla97jpcm}  Of these processes, the
homogenous and heterogeneous growth of ice from liquid water has attracted
considerable attention due to its relevance in atmospheric
physics,~\cite{petrenkobook,koop00nature,bertman05nature,koop05msoc,cantrell05bams}
cryobiology,~\cite{mazur63jgenphys,mazur70science} and in the antifreeze and
food preservation
industries.~\cite{ewart95bioadv,feeney98food,guzman11food,goff12food}  
However, while the thermodynamics of freezing is largely
understood,~\cite{libbrecht05rep,bartels12revmodphys}  
the molecular details of the freezing process are less well established.

Owing to the constantly evolving nature of ice growth, it is difficult to
probe the moving solid-liquid interfacial region experimentally at the
microscopic level.  Measurements have shown the ice-water interface
to be on the order of $\sim 1$~nm wide, or three water layers 
thick, near equilibrium at the melting temperature.~\cite{beaglehole93jpc}
Experiments observing dendritic ice growth have measured maximum growth rates
on the order of $10$~cm/s for the basal plane at temperatures 
$\Delta T_{{\rm M}} = -18$~K below the melting
point.~\cite{puppacher67jcp,langer78jcrys,furukawa93jcrys}   
Experimental deviations of growth rates from theoretical
predictions~\cite{shibkov05jcrys} were
suggested to be due to an unaccounted-for competition between collective
molecular attachment (freezing) and detachment (melting) processes near the solid
surface.~\cite{wilson00philo,frenkel32phys,kolmogorov37ussr, 
johnson40metal,avrami40jcp,shibkov05jcrys}

Molecular dynamics (MD) simulations have proved a useful tool for probing the 
microscopic properties of the (moving) solid-liquid interface more 
directly.~\cite{haymet87cpl,haymet88jcp,nada94jpn,baez95jcp,haymet01jcp, 
haymet02pccp,nada05jcrys,carignano05molphys,kim08jcp,kim09jpca,
kusalik10jacs,rozmanov11pccp,kusalik11crysgrowth,kim12jcp,
rozmanov12pccp,rozmanov12jcp,molinero11nature,molinero12jpcc}  
These studies have also shown the interfacial region is about
three water layers wide, and consists of a slushy mix of ice and liquid features
whose dynamical properties are greatly arrested compared to the
bulk liquid.~\cite{haymet01jcp,carignano05molphys}
Ice growth rates were shown to reach a maximum deep within the supercooled
regime, initially increasing as the temperature was lowered  
below the melting point, but then decreasing upon further undercooling below a
characteristic temperature.~\cite{rozmanov11pccp,weiss11jcp,molinero11nature}
Ideally, one would like to establish how the structure, shape, dynamics, and
molecular attachment rates at the ice surface change with external conditions 
near this crossover point.  If the self-assembly mechanism is a competition between the
rates of attachment and detachment at the liquid-solid contact, what
microscopic properties at the interface favor molecular retention or loss?
Furthermore, what microscopic properties explain why the freezing rate reaches
a maximum in the supercooled regime? 

Here, we use molecular dynamics simulations to investigate how temperature
affects the attachment kinetics of water to the secondary prismatic face 
of ice I$_{{\rm h}}$.  We find the temperature dependence of the ice growth 
rate reaches a maximum when the microscopic efficiency
of converting interfacial water to ice is maximum.  This efficiency
is limited at higher temperatures due to repeated melting and surface
migration~\cite{rost03prl} events across the interfacial regions.
At lower temperatures, the interplay between the roughening of the ice surface 
and increased tetrahedrality of the liquid~\cite{molinero11nature} 
play important limiting roles.
We find molecules which make contact with the intermediate ice layer in
concave regions are more likely to freeze directly.
Molecules that make contact with regions of higher curvature tend to escape
back into the liquid.  Consequently, water molecules are unable to rearrange 
and fit into to increasingly deeper surface pockets at very low temperatures.  
Our results highlight the important role played by  
interfacial water properties in determining the rate of heterogeneous ice 
growth at increasingly larger undercoolings.

\section{Simulations Details} \label{sec:methods}

\begin{figure}[t]
\begin{center}
\includegraphics[width=2.8in]{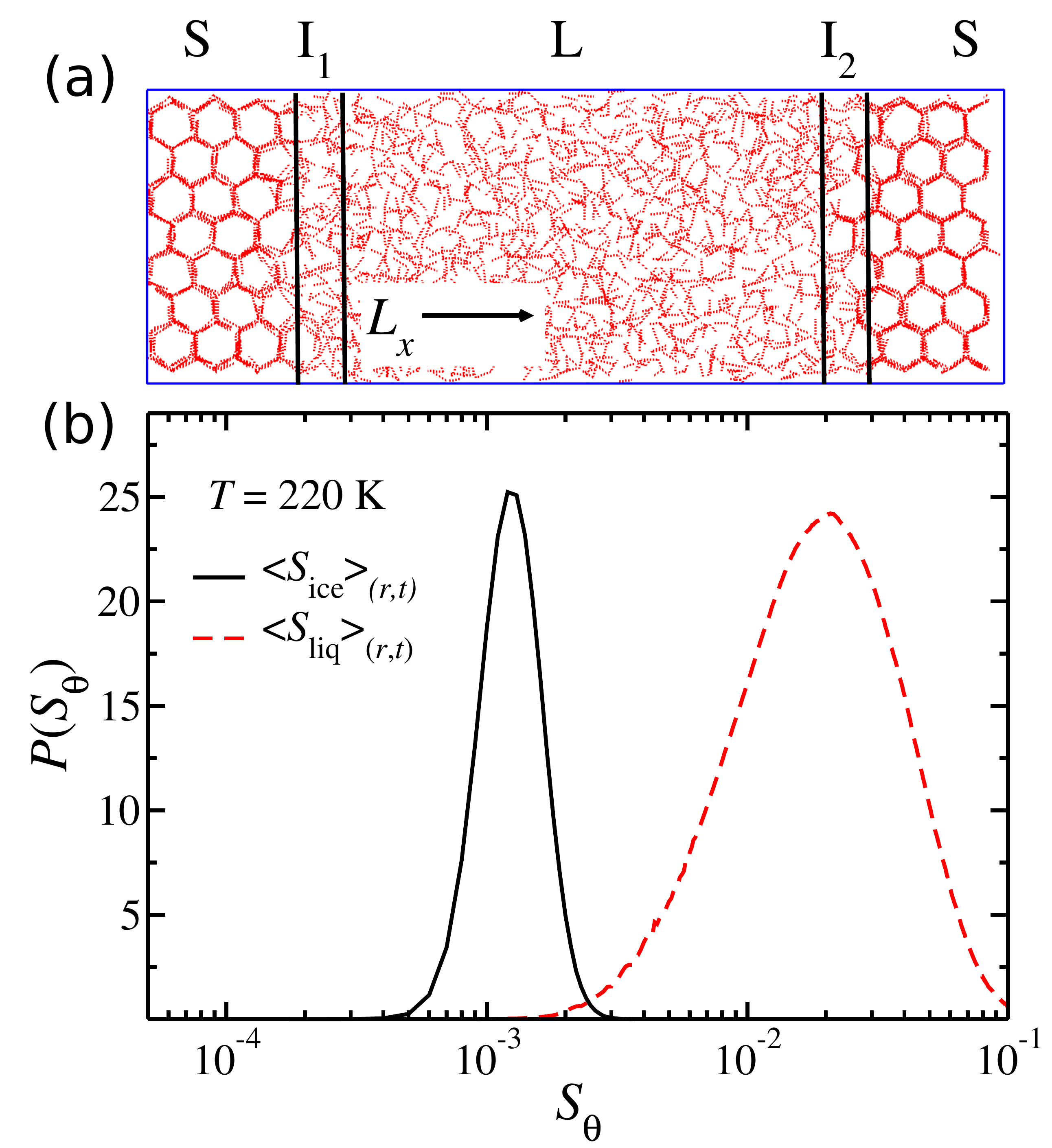}
\caption{\label{fig:sys} 
  (a) Initial configuration used in the simulations with schematic outlines
      of the two interfacial regions (I$_1$, I$_2$) separating the solid (S) 
      and liquid (L) phases. Ice grows along the $L_x$-direction.  
  (b) Tetrahedral order parameter distributions where angle brackets 
      $\langle S \rangle_{(r,t)}$ denote both spatial averaging 
      over several configurations of the trajectory, as well as
      time-based (exponential) smoothing of the resultant instantaneous
      order parameters for each molecule in separate bulk ice and liquid 
      simulations.  Overlap between ice and liquid 
      distributions is significantly reduced using this scheme.
      The ice distribution is normalized by $1/45$ to make the y-axis
      scale more tractable for viewing.}
\end{center}
\end{figure}

Molecular dynamics simulations of the TIP4P/2005 water model~\cite{vega05jcp}
were performed using the GROMACS\cite{gromacs4} package on the ice-liquid
system shown in Fig.~\ref{fig:sys}a under isobaric-isothermal
conditions. Periodic boundary conditions were employed with the long range
electrostatics treated using particle-mesh Ewald summation.~\cite{darden95jcp}
The pressure was kept at 1~bar
using an anisotropic Parrinello-Rahman~\cite{parrinellorahman} barostat with a
time constants of 10~ps.  Constant temperature conditions were imposed using a
Langevin thermostat with a time constant of 4.0~ps, 
which quickly removes latent heat from the
system.~\cite{weiss11jcp,rozmanov11pccp}
Thermostat couplings from 0.1 to 100.0~ps did not affect the
observed freezing rates at $240$~K within statistical error.

Water intermolecular interactions were modeled using
the fully-atomistic TIP4P/2005 potential.  
This model was chosen since it has been shown to accurately reproduce
the high density phase diagram of water and ice,~\cite{vega05jcp} 
as well as the dynamics of water in the supercooled regime.~\cite{stirnemann12jcp}
The melting temperature of the model is $250$~K,~\cite{vega06jcp} which is
considerably more accurate compared with other simple point charge water potentials
given the other advantages of this model.
Freezing rates obtained with the model are also close to experimentally
observed values.~\cite{rozmanov11pccp}

The initial ice structure was prepared according to the Bernal-Fowler
rules,~\cite{bernalfowler33jcp} and brought in contact with an amorphous water
configuration approximately three times the thickness of the ice region. 
The secondary prismatic plane of ice was chosen to contact the water
region since it is the fastest growing face of ice.~\cite{rozmanov11pccp}
The resulting configuration consisted of a sheet of ice (S) and bulk liquid 
(L) separated by two ice-liquid interfacial regions (I$_1$, I$_2$) shown
schematically by the vertical lines in Fig.~\ref{fig:sys}a.  Ice growth was
monitored perpendicular to the interface along the $L_{x}$ direction. 
Two system sizes were used to assess finite size effects.  Ten
trajectories at each temperature were performed using a {\it small}
simulation cell containing 2696 water molecules with approximate dimensions
9.1~nm $\times$ 3.1~nm $\times$ 2.9~nm.  Additionally, three trajectories were 
performed at each temperature using a {\it large} simulation cell, which had
nine times the cross-sectional surface area of the small system and contained
24,264 water molecules.
Data was only gathered until each trajectory was 60\% frozen. This ensured
that the close proximity of the two interfacial regions in the simulations
cell did not affect the analysis at longer times.  

\section{Interface identification}\label{sec:xface}

We developed a robust scheme for classifying the evolving ice, liquid, and
interfacial regions throughout the freezing process.  This classification was
accomplished by employing a suitable order parameter to distinguish between
local ice- and liquid-like structure, and then using profile functions of
this order parameter to identify the instantaneous interface as described below. 

\subsection{Instantaneous molecule classification}

We used a local tetrahedral order parameter to classify whether the local
structure about a water molecule was ice- or liquid-like,~\cite{hardwick98molphys}
\begin{equation}
\label{eqn:tet}
S_{\theta} = \frac{3}{32}\sum_{j=1}^{3} \sum_{k=j+1}^{4} 
            \left( \cos\theta_{jk} + \frac{1}{3}\right)^2 \,
\end{equation}
where the summations extended over all hydrogen bond angles $\theta_{jk}$ defined
by the nearest four neighboring oxygen atoms around a given molecule. 
To enhance our ability to distinguish between ice- and liquid-like
distributions (which can overlap up to 20\% at low temperatures), we used a
combination of position averaging and exponential time smoothing of the
trajectories as described in Appendix~\ref{app:sgt}.  A water molecule was
labeled as ice-like when its resultant order parameter
$\langle S_{\theta} \rangle_t$ was less than a carefully selected threshold
criteria $S_{{\rm tol}}$.  Using the smoothing procedure, near perfect separation
between ice and liquid configurations can be obtained with less than 4\%
overlap between bulk ice and bulk liquid distributions at $220$~K, as shown in
Fig.~\ref{fig:sys}b.  The critical threshold value were chosen to be 
$S_{{\rm tol}} = 0.002$ for $220$~K, and $S_{{\rm tol}} = 0.005$ for all other
temperatures. At $300$~K, no water molecules were misclassified as ice-like in
simulations of the bulk liquid.  All other observables were calculated
from the raw unaveraged trajectories.
\begin{figure}[t]
\begin{center}
\includegraphics[width=2.8in]{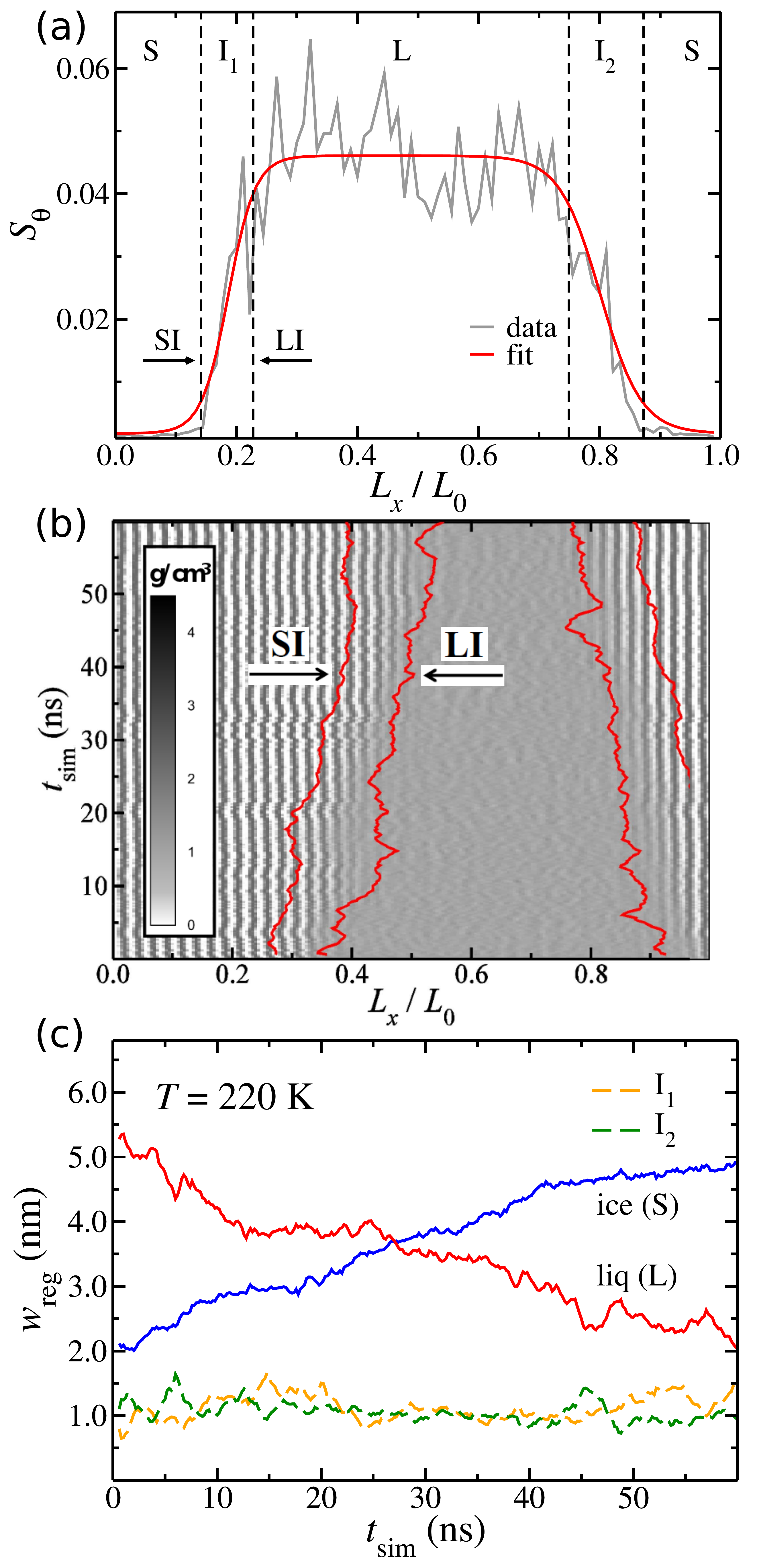}
\caption{\label{fig:xface}
  (a) Tetrahedral order parameter binned along direction of ice growth
      and resultant profile function used to identify the solid-interface
      (SI) and liquid-interface (LI) dividing surfaces.
  (b) Contour map of the density profile evolving during a typical 
      trajectory with the solid lines giving the instantaneous 
      interfacial boundaries (SI and LI) for each interfacial region.  
  (c) Average resultant thicknesses of the ice, liquid, and the 
      widths $w_{{\rm reg}}$ of the interfacial regions (I$_1$, I$_2$).}
\end{center}
\end{figure}

\subsection{Instantaneous interface classification}
\label{sec:int_class}

The instantaneous positions of the interfacial regions were identified from
profile functions of the tetrahedral order parameter $\langle S_{\theta} \rangle_t$ 
projected along the direction of ice growth.  This approach is similar to
methods used in previous studies to define the extent of the interfacial 
region.~\cite{haymet87cpl,haymet01jcp,razul11jcp}
Here, we track two interface boundaries as shown in Fig.~\ref{fig:xface}a: 
the solid-interface (SI) and the liquid-interface (LI).  These dividing surfaces
(dashed vertical lines in the figure) were identified using the procedure
described in Appendix~\ref{app:xface}. 

Fig.~\ref{fig:xface}b shows the projected
density along the scaled simulation box for a typical trajectory at 
$220$~K obtained using this scheme. From these boundary lines, the widths of 
the interfacial regions (I$_1$, I$_2$), and the thicknesses of the ice (S) 
and liquid (L) regions can be monitored as freezing occurs, as shown in
Fig.~\ref{fig:xface}c. Importantly, the interfacial widths remain roughly
constant throughout the simulation, and are on the order of $\sim 1$~nm, 
or approximately three water layers, consistent with experimental 
measurements~\cite{beaglehole93jpc} and other MD 
studies.~\cite{haymet01jcp,razul11jcp,kusalik11crysgrowth,molinero12jpcc}

Since ice growth does not proceed on a layer-by-layer basis along the prismatic
directions,~\cite{nada94jpn,carignano05molphys,kim12jcp} we modeled the
rough SI and LI surfaces by dividing the cross-sectional area of the
simulation box into a 2D array of fibers extending the entire length of the
system along the direction of ice growth.  Each square fiber was 
defined by a feature size of $d_{{\rm f}} = 0.6$~nm sides.  
The envelope functions, and subsequently the positions of the SI and LI 
boundaries, were formed separately in each fiber.

\section{Results and discussion} \label{sec:results}

\subsection{Growth rate maximum}

\begin{figure}[t]
\begin{center}
\includegraphics[width=2.8in]{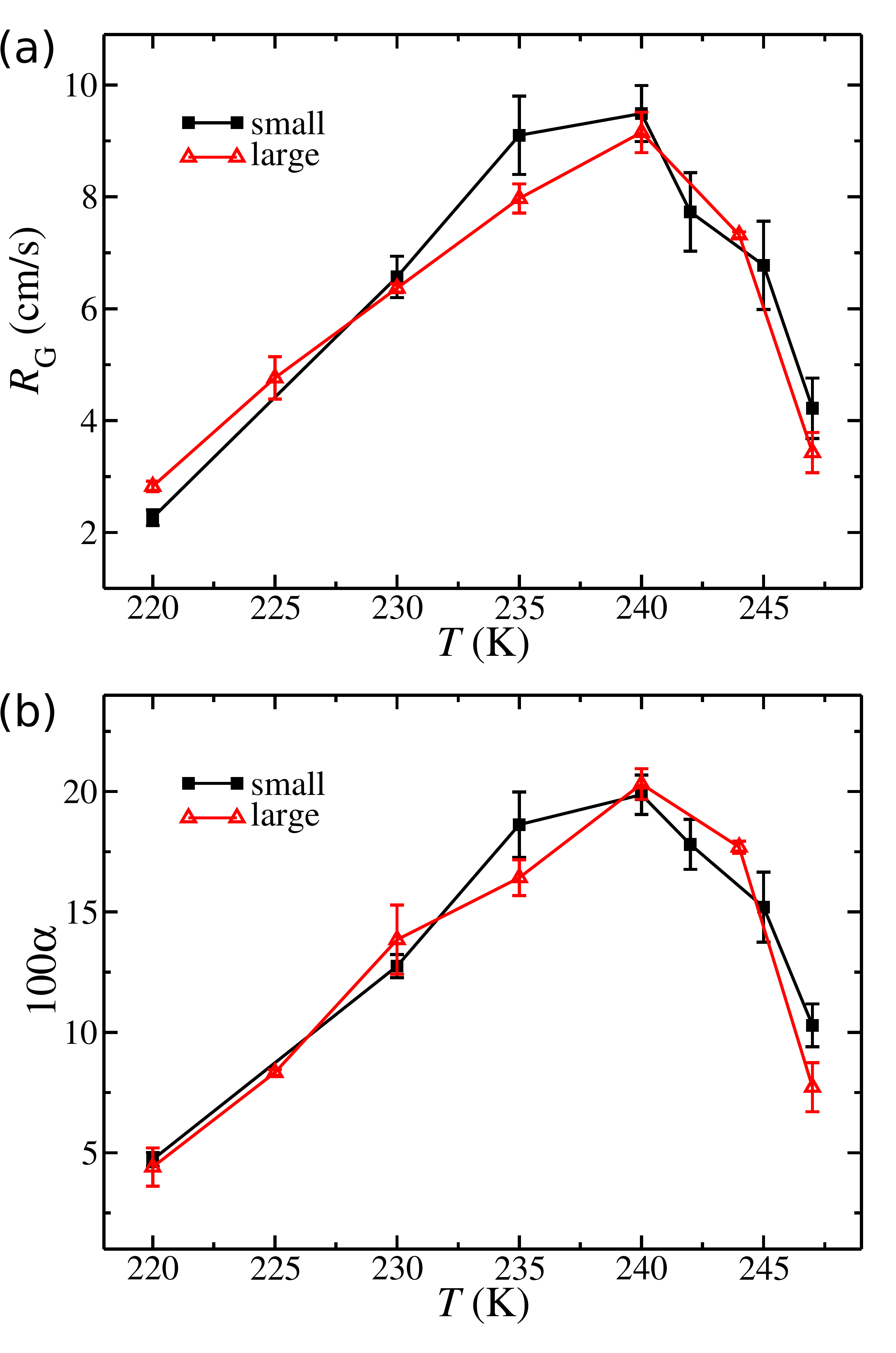}
\caption{\label{fig:freezerates_alpha}
    (a) Average crystal growth velocities $R_{{\rm G}}$ for the secondary
        prismatic face of ice I$_{{\rm h}}$ for the two system sizes. 
    (b) The retention probability $\alpha$ defining the system efficiency of
        converting liquid water molecules to ice.}
\end{center}
\end{figure}

The measured ice growth rates $R_{{\rm G}}$ along the secondary
prismatic direction for different temperatures is shown in
Fig.~\ref{fig:freezerates_alpha}a.
The growth rate profile reaches a maximum of $9.6 \pm 0.5$ cm/s
at $240$~K ($\Delta T_{{\rm M}} = -10$~K), in good agreement with previous
results using this water model.~\cite{rozmanov12jcp} 
Experimental studies of the growth of ice dendrites
report growth velocities of 10-12 cm/s for the fastest
growing ice faces at temperatures $\Delta T_{{\rm M}} = -18$~K below the melting
point.~\cite{puppacher67jcp,bauerecker08jpcc} 

The maximum in the growth rate in Fig.~\ref{fig:freezerates_alpha}a is
characteristic of a crossover from thermodynamically-driven to kinetic-limited 
crystal growth.~\cite{kirkpatrick75mineral,molinero11nature,rozmanov11pccp,weiss11jcp}
From $247$ to $240$~K, the rate of crystal growth increases by a
factor of $\sim 2$ as the chemical potential difference between the
liquid and solid phases increases (i.e. $\mu_{{\rm L}}-\mu_{{\rm S}}>0$)
and ice becomes thermodynamically more
favorable.~\cite{bartels12revmodphys}
Below $240$~K, the growth rate progressively decreases and is a 
factor of $\sim 5$ smaller at $220$~K than the maximum.
This decrease is described as arising from increasing kinetic
barriers governing activated processes such as diffusion,
which become rate-limiting below $240$~K.~\cite{kirkpatrick75mineral,rozmanov11pccp}   
We note the lowest temperature $220$~K is below where the peak in the isobaric
heat capacity occurs for this water model ($\sim 225$~K -- which signifies the
onset of water's so called no-man's 
land),~\cite{stanley98nature,molinero10jcp,vega10jcp,carignano13jcp}  
and was included to see if the scaling of the growth process continues as the
system enters this deeply cooled regime.

To assess the efficiency with which water molecules in the interfacial
regions are incorporated into the ice phase at the varying temperatures, 
we consider the interfacial retention probability $\alpha$, 
\begin{equation}
\label{eqn:alpha}
\alpha = \frac{\Phi_{{\rm F}}}{\Phi_{{\rm F}} + \Phi_{\rm E}}\,,
\end{equation}
where $\Phi_{{\rm F}}$ is the flux (number of molecules per nm$^2$ per ns) of
liquid molecules that irreversibly freeze to the solid surface, 
and $\Phi_{{\rm E}}$ is the flux of liquid molecules that enter the
interfacial region but later escape to the liquid without freezing.
Written in this way, $\alpha$ is the liquid-to-ice conversion efficiency
of the system, or the percentage of water molecules that 
freeze from the total number of water molecules that cross into the
interfacial regions (I$_1$ and I$_2$) from the liquid.

The temperature dependence of the retention probability $\alpha$ in
Fig.~\ref{fig:freezerates_alpha}b follows the same trend observed in the
growth rate profile, exhibiting a maximum at $240$~K and minima at the
lowest and highest temperatures $220$ and $247$~K for both system sizes studied.
This temperature dependence is intuitively expected 
since the crystal growth velocity is microscopically determined by the
rates which molecules become incorporated into the solid surface.
Crystal growth will be limited if the conversion of liquid molecules to ice
is low, as is the case at $220$ and $247$~K in
Fig.~\ref{fig:freezerates_alpha}b.

\begin{figure}[t]
\begin{center}
\includegraphics[width=2.8in]{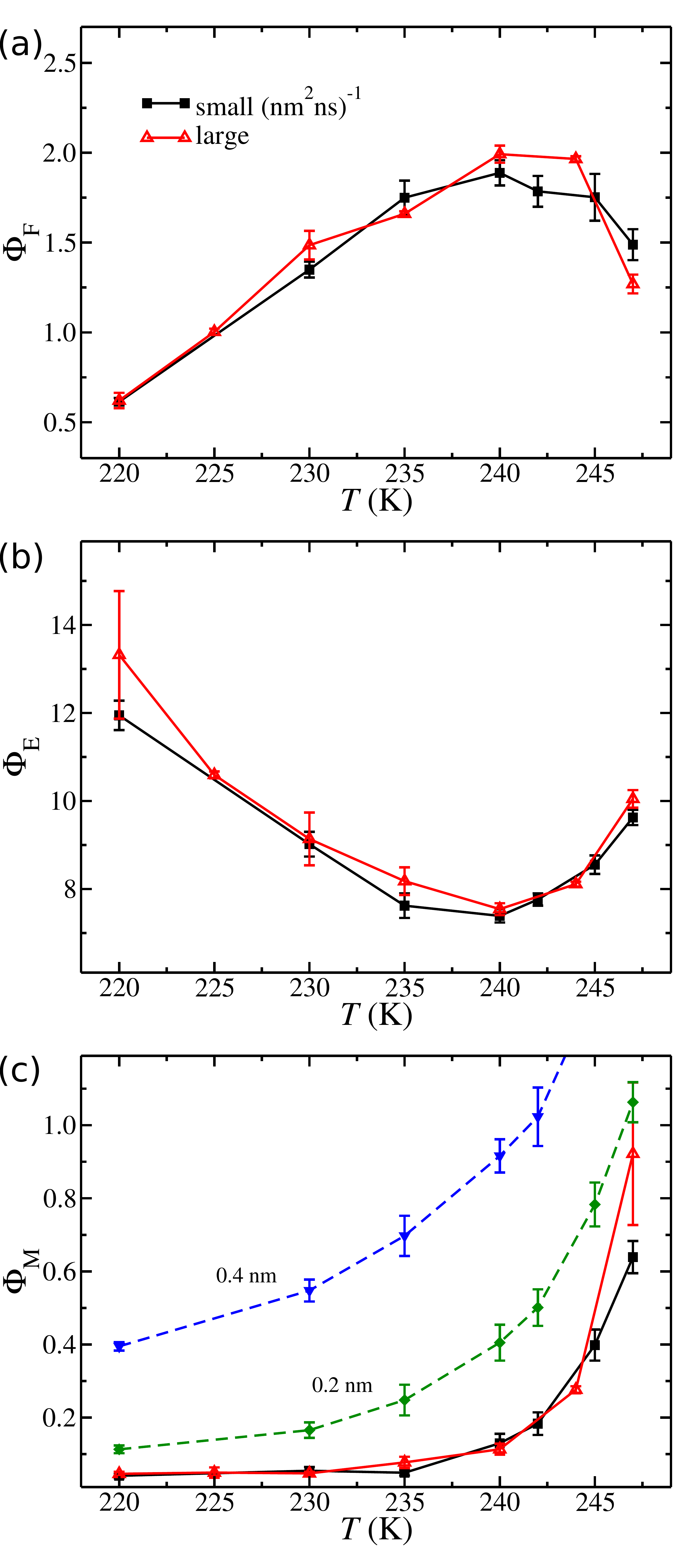}
\caption{\label{fig:flux}
    (a) The flux $\Phi_{{\rm F}}$ of molecules that irreversibly freeze to the
        solid phase.
    (b) The flux $\Phi_{{\rm E}}$ of molecules that cross into the interfacial
        regions but later escape to the liquid without contacting the ice surface.
    (c) The flux $\Phi_{{\rm M}}$ of molecules that escape to the liquid
        (melt) from the ice surface (solid lines), and 
        ice-like planes progressively farther into the
        interfacial region (dashed lines).}
\end{center}
\end{figure}

The individual contributions to the retention probability in
Fig.~\ref{fig:flux} allow the origins of the freezing efficiency to be 
assessed.   As can be seen from Fig.~\ref{fig:flux}a, the flux of molecules 
that irreversibly freeze $\Phi_{\rm F}$ increases by nearly a factor of 
$\sim 2$ as the temperature is lowered from $247-240$~K and 
ice becomes thermodynamically more favorable.
There are fewer irreversible freeze events at $247$~K because of an increased
propensity to melt, or to detach from the ice-like layers near the solid
surface (see next paragraph), which is intuitively expected near the melting point.
Below $240$~K, however, $\Phi_{\rm F}$ begins to 
decrease significantly from its maximum value, 
by up to a factor of $\sim 4$ at $220$~K 
(a similar change in freezing rate is observed in 
Fig.~\ref{fig:freezerates_alpha}a).
While the increased tetrahedrality of the liquid~\cite{molinero11nature}
and corresponding slow-down in water dynamics~\cite{weiss11jcp} are expected to
contribute to the observed decrease in freeze events, we examine further below
how changes in the interfacial structure can also play a limiting role on 
the attachment kinetics at low temperatures.

The escape flux of molecules that enter the interfacial region but 
return to the liquid is shown in Fig.~\ref{fig:flux}b.  
The nonmonotonic behavior shows that the number of escape events is
(somewhat unintuitively) greater at both $220$ and $247$~K.
An increased propensity to detach from the ice
surface (or ice-like layers near the ice surface), plays an increasingly
greater contribution to the escape flux as the temperature increases.
Fig.~\ref{fig:flux}c shows how the flux $\Phi_{{\rm M}}$ of frozen molecules 
that detach from the ice and ice-like planes near the solid surface, i.e. melt
events, increases with temperature and proximity to the liquid region (dashed
lines), as is intuitively expected.
At temperatures below $240$~K, however, increasing values for $\Phi_{{\rm E}}$
are more puzzling. This greater propensity to return to the liquid will be
attributed to a roughening of the intermediate ice layers, as will be
discussed further below.

The microscopic population analysis shows that although the system has
similarly low retention probability and ice growth rates at $220$ and $247$~K,
the reasons for the limited growth velocities are quite different.  At high
temperatures, the growth rate is limited due to an increased melting
propensity.  At low temperature, the growth rate is limited by a
sharp decrease in the number of direct freezing events due to an inability to convert
interfacial water molecules into ice before these molecules too escape back to the
liquid.  In the following sections, we analyze how the microscopic dynamics
and interface topology contribute to these observations. 

\subsection{Effects of interface topology}

It is constructive to analyze the collective freezing and melting
events at the solid-interface (SI) guided by a simple model.
As discussed in Sec.~\ref{sec:int_class}, we approximate the roughening
of the interfacial dividing surfaces by fitting envelope functions of
$\langle S_{\theta} \rangle_t$ in a series of discretized rectangular
fibers extending the length of the simulation box.
If the fluctuating position of the solid-interface in
each fiber is treated as a biased random walker,
where a step forwards represents a freezing
event, and a step backwards represents a melting event,
and where the random walk is biased by the degree of undercooling, then
ensemble averages over all the fibers derived using the moment generating 
function for continuous walks will satisfy,~\cite{spitzerbook}
\begin{equation}
\label{eqn:kvel}
\langle \delta x(\tau) \rangle = \langle x_i(t_{{\rm 0}}+\tau) 
    - x_i(t_{{\rm 0}}) \rangle = 
\left( k_{{\rm f}} - k_{{\rm b}} \right)a \tau
\end{equation}
and
\begin{equation}
\label{eqn:kdisp}
\langle \delta x^2(\tau) \rangle 
= \langle \delta x(\tau)^2 \rangle 
- \langle \delta x(\tau) \rangle^2 
=
\left( k_{{\rm f}} + k_{{\rm b}} \right) a^2 \tau
\end{equation}
where $x_i(\tau)$ is the position of the $i$'th cell or fiber at time
$\tau$, $k_{{\rm f}}a$ and $k_{{\rm b}}a$ are the forward (freezing) and
backward (melting) rates, and $a$ is the average measured step
size.  Eqn.~\ref{eqn:kvel} gives the growth rate (or velocity) as 
$R_{\rm G}=(k_{\rm f}-k_{\rm b})a$.  Eqn.~\ref{eqn:kdisp} is a measure of the
spread of the random walker trajectories.  Using these relations, the forward
and backward rates can be extracted for each temperature.  

\begin{figure}[t]
\begin{center}
\includegraphics[width=2.8in]{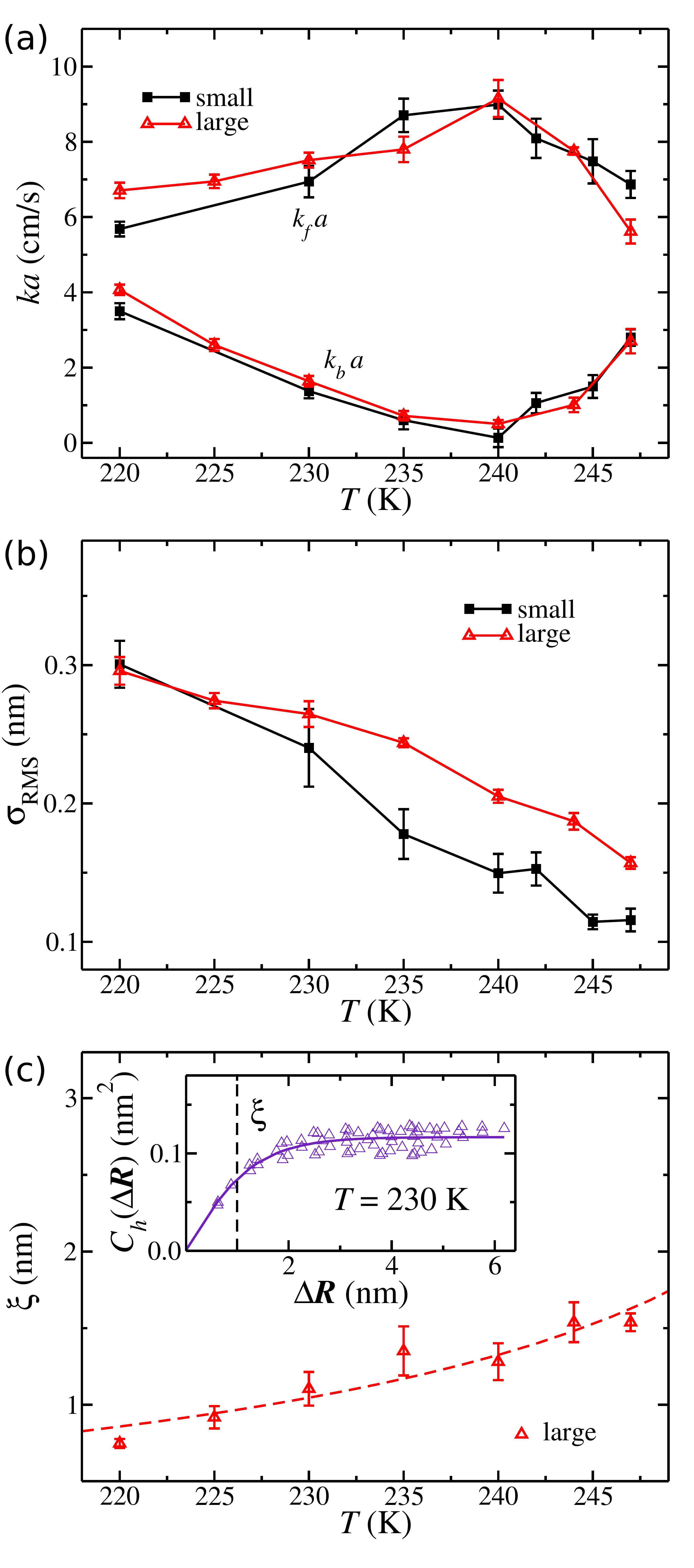}
\caption{\label{fig:kdata}
        (a) Forward ($k_f$) and backward ($k_b$) rates
            representing freezing and melting processes of the 
            solid-interface for the two different sized systems.
            Results are multiplied by the measured average step size
            $a=0.23$~nm which roughly equals the spacing between 
            successive ice planes. 
        (b) The root-mean-squared deviation of the discretized roughened
            ice surface $\sigma_{{\rm RMS}}$.  
        (c) Spatial correlation lengths $\xi$ for the large
            systems extracted by fitting the height-difference
            correlation function shown in the inset as an example at 
            $T = 230$~K.
        }
\end{center}
\end{figure}

Fig.~\ref{fig:kdata}a shows the nonmonotonic temperature dependence of the 
forward ($k_f$) and backward ($k_b$) rates for the two system sizes using a
feature size of $d_{{\rm f}} = 0.6$~nm for the square fibers defining the
discretized random walkers.  We have separately verified that the trends are
qualitatively reproduced using:  (1) Thicker fibers up to $d_{{\rm f}} =
1.0$~nm, (2) using interfacial profile functions constructed from the total
density $\rho$,~\cite{haymet01jcp} and (3) using trajectories generated using
the standard TIP4P water model~\cite{tip4p83jcp} (at relative undercoolings). 

The forward rate $k_f$ increases by a factor of $\sim 1.3$ from $247-240$~K,
and progressively decreases by a factor of $\sim 1.5$ from $240-220$~K. The
backward rate shows a stronger nonmonotonic temperature dependence:
decreasing by a factor of $\sim 5$ from 247~K to the minimum at $240$~K, and
increasing by roughly the same factor upon further cooling to $220$~K.  
The nonmonotonic behavior in Fig.~\ref{fig:kdata}a shows that the difference
between forward and backward processes, and consequently the growth rate
$R_{{\rm G}} = (k_f - k_b)a$, is greatest at $240$~K and lowest at both $220$
and $247$~K.  While low thermodynamic driving force and increased melting
propensity can intuitively describe the small difference in $(k_f - k_b)a$ at
$247$~K (see also Figs.~\ref{fig:flux}a-b), it is instructive to analyze how
changes in interfacial structure coupled with arrested water
dynamics~\cite{weiss11jcp} and increased tetrahedrality of the
liquid~\cite{molinero11nature} can limit the attachment kinetics that
microscopically underpin the forward and backward rates at low temperatures.

The microscopic features of the interfacial regions are intricately
linked to nonmonotonic temperature dependence of the forward and
backward processes at increasingly lower temperatures.  Notably,
the structural characteristics of the solid-liquid interface scale
differently than in solid-vapor systems.  
Fig.~\ref{fig:kdata}b shows the root-mean-squared (RMS) deviation 
of the discretized fibers used to define the instantaneously
roughened solid-interface for all temperatures.  
The figure shows that the dividing surface becomes
rougher as the temperature is lowered, consistent
with previous simulations,~\cite{kusalik11crysgrowth} but
opposite to what is observed at the ice-vapor
interface.~\cite{hilliard58jcp,libbrecht05rep}
The RMS deviation is slightly larger than 0.3~nm at the coldest temperatures,
corresponding approximately to an extra ice layer, or half a hexagon as viewed
from the basal direction.  The larger systems show higher RMS deviations
at higher temperatures, which is a notable finite-size effect.

Small systems in particular can lead to overestimated growth 
kinetics.~\cite{rozmanov11pccp}  To understand how this arises,
Fig.~\ref{fig:kdata}c shows the roughness correlation 
length $\xi$ extracted from the height-difference spatial correlation
function for the large systems,~\cite{weeks79acp,salditt95prb} 
\begin{equation}
C_h(\Delta {\bf R}) = \langle | h({\bf r}) - h({\bf r'}) |^2 \rangle \,,
\end{equation}
where $h({\bf r})$ is the height of the discretized fiber 
(in this case, its extent along the $x$-direction) 
at position ${\bf r}$ along the cross-sectional 
area of the roughened surface, and $\Delta {\bf R} = {\bf r}-{\bf r'}$.
The roughness correlation length was extracted by fitting
the function $C_h(\Delta {\bf R})$ to the form,~\cite{salditt95prb}
\begin{equation}
C_h(\Delta {\bf R}) = 2\sigma^2\left\{ 1-\exp\left[ 
                  - (\Delta {\bf R}/\xi)^{(2P)}\right] \right\} \,,
\end{equation}
where $\sigma$ is the standard deviation in heights, $\xi$ is the
correlation length, and $P$ is a roughness exponent, typically 
between 0.5 and 0.6 for the large systems.
The temperature dependence of $\xi$ can be fitted to the Kosterlitz-Thouless 
scaling relation (dashed line), which increases exponentially and diverges at
the roughening transition temperature near the melting point.~\cite{kosterlitz73jpc} 
At $247$~K, the roughness correlation length is nearly $1.6$~nm, almost half
the size of the large system simulation cell.
These data indicate that finite size effects will become prevalent if 
long-wavelength capillary waves driving the roughening transition are 
damped out due to small simulation cells.  Higher observed growth 
rates in sufficiently small systems could result due to the appearance 
of defected surface motifs that aid molecular rearrangement near the 
solid surface.~\cite{weeks79acp,broughton83jcp}

The increase in the roughness correlation length with 
temperature in Fig.~\ref{fig:kdata}c shows
structural features are more correlated at higher 
temperatures.  Consequently, the solid-interface appears smoother, with a
smaller RMS deviation in the height profiles of the discretized fibers.  
At lower temperatures, structural correlations occur over shorter
lengthscales, and the ice surface becomes rougher.  

To gain insight into which properties of the roughened interface 
inhibit molecular retention at low temperatures, we track where
incoming molecules contact the intermediate ice layer 
(IIL), which is typically composed of a few roughened water 
layers above the ice boundary as described in Appendix~\ref{app:sgt}.
We divide the flux of these incoming molecules into molecules that freeze, 
and molecules that escape back to the liquid without being incorporated in 
the solid phase.  We measure the average distance $d$ of a water molecule 
to this IIL, and additionally, the curvature $\kappa$ of the roughened surface 
at the contact point.  Contact is defined by the formation of hydrogen bonds 
between the incoming water molecule and molecules that are part of the IIL.
The mean curvature $\kappa$ at each point of the atomistically rough intermediate
ice surface was calculated from the average of the principal curvatures, 
$\kappa = \frac{1}{2}(\kappa_{{\rm min}}+\kappa_{{\rm max}})$, 
where $\kappa_{{\rm min}}$ and $\kappa_{{\rm max}}$ are the eigenvalues
of the shape operator representing the minimum and maximum degree of
deflection of a surface at a given point.~\cite{toponogovbook}
\begin{figure}[t]
\begin{center}
\includegraphics[width=2.8in]{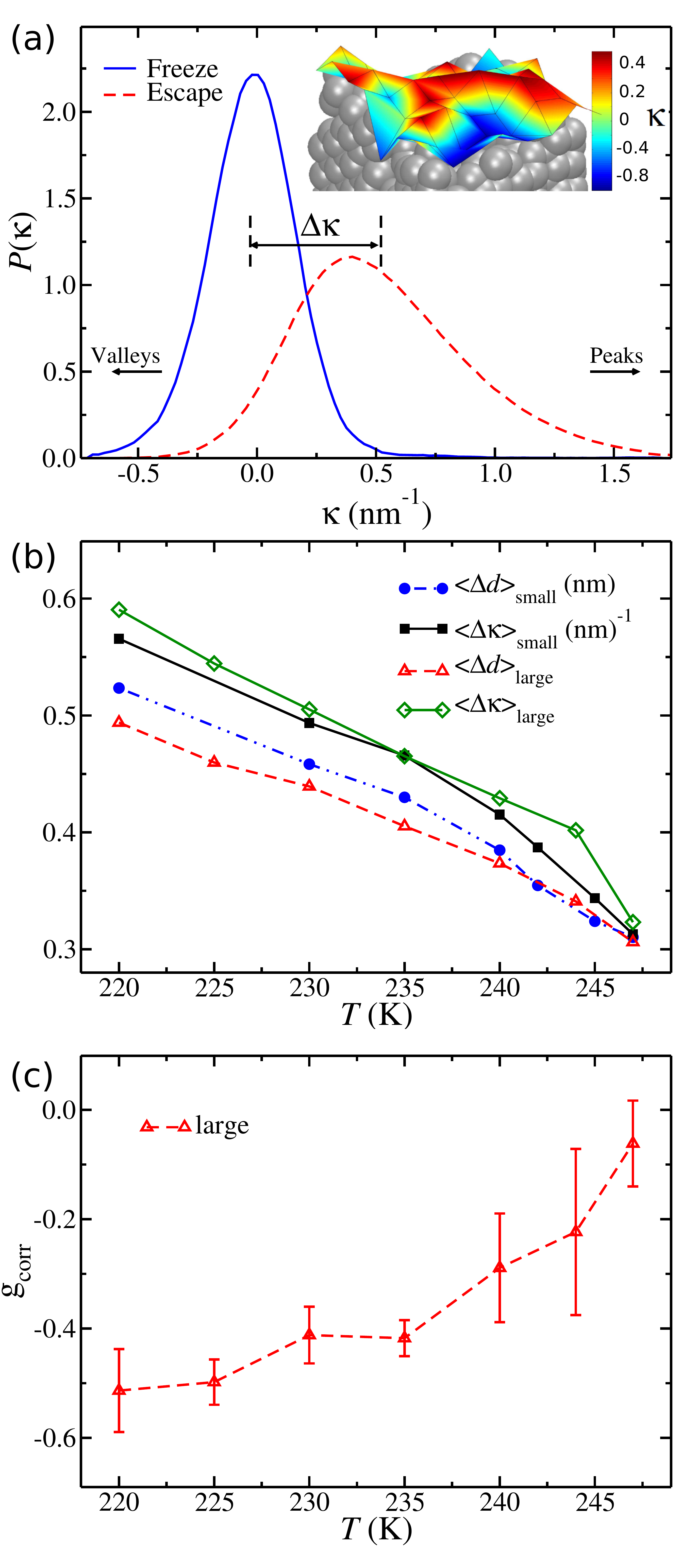}
\caption{\label{fig:curv}
        (a) Mean curvature distributions $\kappa$ of freeze,
            and escape populations at $T = 230$~K.  The inset
            shows a colormap of the resulting curvature values atop an
            instantaneous configuration of the intermediate ice layer.
        (b) Differences between freeze and escape distributions
            for the two measures, $\langle \Delta d \rangle$ and
            $\langle \Delta \kappa \rangle$, for the two system sizes.
        (c) Statistical correlation $g_{{\rm corr}}$ between low growth
            rate and high surface roughness for the large systems.
        }
\end{center}
\end{figure}

As an example, Fig.~\ref{fig:curv}a shows the curvature distributions
for the freezing and escaping populations averaged over all
trajectories at $230$~K.  Molecules that directly freeze tend to dock
in concave valleys, or regions of negative (or near zero) mean
curvature on the roughened intermediate ice surface.  Conversely,
molecules that escape back into the liquid tend to bind to peaks, or
regions of high mean curvature farther away from the solid interface.
Particles subsequently have a greater propensity to escape from convex
surfaces than from concave surfaces of ice, as seems to be the case
for liquid-vapor interfaces as well (see
Refs.[\onlinecite{weeks79acp,chandler10jpcb}]).

Fig.~\ref{fig:curv}b measures the degree of overlap 
between the freeze and escape distributions for both observables, 
quantified by taking differences in the mean values:
$\Delta \kappa = \langle \kappa_{{\rm escape}} \rangle - 
\langle \kappa_{{\rm freeze}} \rangle$ and
$\Delta d = \langle d_{{\rm escape}} \rangle - 
\langle d_{{\rm freeze}} \rangle$.
The differences in both $\Delta \kappa$ and $\Delta d$ decrease as
the temperature approaches the melting point, where the
solid surface is much flatter.  Analogously, there are fewer 
peaks and valleys for the smoother interfaces at higher temperatures,
and less of a docking preference between freeze and escape populations.
At large undercoolings, however, the separation between distributions 
is larger, indicating that escape events preferentially bind to regions of
higher curvature.  
These roughened structural features may develop in order to expose the
more stable prismatic face to the liquid
contact.~\cite{nada05jcrys,carignano09jpcc} 
Although, surface roughening has been noted to appear on the prismatic and
basal faces of ice as well.~\cite{kim12jcp}

To asses how the roughened structural features of the ice surface
impact the growth kinetics, Fig.~\ref{fig:curv}c shows Pearson's
statistical correlation coefficient $g_{{\rm corr}}$ between the instantaneous 
growth rates $R_{{\rm G}}$ and surface roughness $\sigma_{{\rm RMS}}$ for the large 
systems.  Instantaneous growth rates were obtained from 2.0~ns moving windows 
centered at each frame of the trajectories.  The coefficient is $-1.0$ when
growth rates are completely anticorrelated with surface roughness.  
When $g_{{\rm corr}}$ is 0.0, the two measures are uncorrelated.  
As can be seen in the figure, $g_{{\rm corr}}$ approaches 0.0 at high 
temperatures, and $-0.51 \pm 0.08$ as the temperature is 
lowered to 220~K.  Consequently, {\it low} growth rates are increasingly 
associated with {\it high} surface roughness below the temperature 
of maximum crystallization.  

\section{Summary and conclusions} \label{sec:conclusions}

Using molecular dynamics simulations, we have identified
structural features of the ice-liquid interface along the secondary prismatic
direction that affect crystal growth velocities in the supercooled regime.  
Near the melting point, the freezing rate is limited by surface depletion
events, as molecules detach from the ice and migrate back to the liquid.  At
much lower temperatures, the topology of the roughened intermediate ice layer
plays an important limiting role hindering ice growth.  The decrease in the 
interfacial retention probability at low temperatures is due to the appearance 
of high-curvature structural motifs.  Along with the increased tetrahedrality
of the liquid,~\cite{molinero11nature} these roughened structural profiles
limit the crystal growth velocities at larger undercoolings, as the liquid is
unable to adjust to the required surface geometry.

At the temperature of maximum crystallization, the efficiency of 
converting interfacial water to ice is maximized.  The rates between 
competing attachment and detachment reactions in the interfacial region 
is greatest at this temperature.  Notably, molecular detachment rates leading
to surface melting are minimized when the crystallization rate is maximized.  
The liquid is best able to adjust and fill surface pockets at the temperature
of maximum growth.

These insights into the molecular scale rate limiting processes for
heterogeneous ice nucleation should prove useful in analyzing how
other perturbations, such as the presence of solutes, affect the
interfacial region and freezing rates.  Such an understanding is vital
for unraveling the ice growth inhibition mechanisms of antifreeze proteins 
in biological systems, and for industrial cryogenics applications.

\section{Acknowledgments} \label{sec:acknowledgments}

The authors would like to thank Valeria Molinero for her 
insightful comments on an early version of this manuscript.  
This research was supported by a grant to BJB from the
National Science Foundation (NSF-CHE-0910943).  
We thank CCNI at RPI and the XSEDE resources at TACC for providing
computational facilities to support this project. 

\appendix 

\section{Ice-water selectivity}
\label{app:sgt}

In order to enhance the selectivity between ice- and liquid-like
local configurations using the tetrahedral order parameter in Eqn.~\ref{eqn:tet},
the atomic positions of the trajectories were first averaged over 5~ps
  windows to reduce thermal and librational noise in the oxygen atom
  positions.  This time is shorter than the $\sim 10$~ps characteristic water
  reorientation time for this water model at 250~K,~\cite{stirnemann12jcp}
  which is a higher than any temperature used here and so is much shorter
  than the characteristic time in which a molecule can interconvert between ice
  and water in our trajectories.  These averaged positions were then used to
  evaluate the tetrahedral order parameter using Eqn.~\ref{eqn:tet} for each
  water molecule.

The order parameter history of each water molecule from the position averaged
  trajectories was then exponentially time-smoothed using, $\langle S_{\theta}
  \rangle_t = \alpha S_{\theta}(t) + (1-\alpha) \langle S_{\theta} \rangle_{t-1}$,
  where $S_{\theta}(t)$ is the instantaneous order parameter of a given
  molecule at time $t$, $\langle S_{\theta} \rangle_{t-1}$ is the smoothed order
  parameter of the previous (position-averaged) time step $t-1$, and $\alpha$
  is the smoothing parameter.  We found $\alpha\sim 0.3$ gave adequate
  separation between bulk ice and liquid distributions at low
  temperatures. This time-based smoothing was used to inhibit instantaneous
  tetrahedral configurations from contributing to the analysis, which may
  spontaneously occur even at high temperatures.

Molecules within the interfacial regions whose order parameter 
$\langle S_{\theta} \rangle_t$ was greater than the ice threshold criteria
$S_{{\rm tol}}$, but less than 75\% of the liquid value at the given
temperature, were labeled as intermediate ice.~\cite{molinero11nature}
A molecule retained its ice, liquid, or intermediate ice label for the
duration of the 5~ps position-averaging window.  These labels were only used 
for population analysis. 

\section{Interface identification}
\label{app:xface}

In order to identify the (moving) positions of the solid- and liquid-interface
dividing surfaces,
the order parameter $\langle S_{\theta} \rangle_t$ was binned across the
  simulation cell as shown in Fig.~\ref{fig:xface}a.  The resulting
  distributions were fitted to profile functions of the form $f(x) = A\left[
  \tanh\left(\frac{x-C}{B}\right)  -\tanh\left(\frac{x-D}{E}\right) \right]$,
  where $\{A,B,C,D,E\}$ were the fit parameters.  The roots of the fourth
  derivative of this function lie close to the positions of the shoulders of
  the profile and were used here to define the locations of the SI and LI
  dividing surfaces as shown by the dashed vertical lines in Fig.~\ref{fig:xface}a.

The distributions were accrued over 200~ps time windows. This sampling
  corresponded to roughly one-tenth the time it took the solid-interface to
  pass through the next layer of ice in the fastest growing trajectories.  
The positions of the SI boundaries were aligned with the nearest ice plane
  at each time step since the lower shoulders of the fitted profile function
  were not necessarily concomitant with the outermost ice layer.
Once the interfacial boundaries were established, molecules could be 
  identified as being in the solid (S), liquid (L), or in one of the two 
  interfacial regions (I$_1$, I$_2$) at any given time in the trajectory.
  To remove rapid recrossing events across the boundaries, a molecule
  was required to reside a minimum of 200~ps in a region before it counted
  towards the population statistics. 
Molecular configurations and surfaces were rendered using MATLAB
and the Visual Molecular Dynamics (VMD) package.~\cite{matlab05,VMD}


\end{document}